\documentclass{article}
\usepackage{spconf,amsmath,graphicx}
\usepackage{amssymb} 
\usepackage{algorithm,algpseudocode} 
\usepackage{makecell}
\usepackage{hyperref}

\usepackage[caption=false,font=footnotesize]{subfig}

\def\L{{\cal L}}

\title{Reduced-complexity Adaptive Loop Filtering via Input-dependent Graph Filters}
\twoauthors
  {Wen-Yang Lu, Eduardo Pavez, Antonio Ortega}
  {University of Southern California\\
   Department of Electrical and Computer Engineering\\
   Los Angeles, CA, USA}
  {Roman Chernyak, Shan Liu}
  {Tencent Media Lab\\
   Palo Alto, CA, USA}

\begin{document}
\ninept
\maketitle
\begin{abstract}
Adaptive Loop Filtering is an important tool for suppressing compression artifacts in modern video codecs. In the enhanced compression model (ECM), a software test model used for experimenting with video coding tools beyond Versatile Video Coding, fixed filters are trained offline and achieve high signal adaptivity via a fine-grained gradient-based classifier, resulting in a large number of fixed filters that introduce redundancy and increased implementation complexity. Reducing this redundancy without compromising artifact suppression, therefore, remains a key challenge. This paper proposes an alternative graph-based fixed-filtering framework for adaptive loop filtering. By using a graph to encode pixel-intensity relationships, our approach captures local structural information more effectively than gradient-based classification alone. Fixed filters are learned as polynomial graph filters, enabling structurally similar local patterns to share common filtering behavior. Experimental results demonstrate that the proposed approach achieves a comparable performance to the ECM baseline while reducing the number of required filters by an order of magnitude.

\end{abstract}
\begin{keywords}
Graph signal processing, adaptive loop filtering, video coding, fixed filters
\end{keywords}
\section{Introduction}
\label{sec:intro}
Modern video coding standards rely on in-loop filtering tools to suppress compression artifacts and improve reconstruction quality.
Among these tools, Adaptive Loop Filtering (ALF) \cite{Tsai2013ALF, Karczewicz2021inloop} plays a critical role by refining reconstructed samples through content-adaptive filtering.
By exploiting local signal characteristics, ALF can effectively reduce artifacts introduced by block-based prediction, transform, and quantization.
In the enhanced compression model (ECM) \cite{coban2025ecm19}, a software test model that has been used for testing video coding tools beyond Versatile Video Coding (VVC) \cite{Bross2021OVVC}, 
ALF incorporates both online adaptive filters and a set of offline trained fixed filters.
These filter outputs are aggregated through an online-learned Wiener filtering process~\cite{Tsai2013ALF} to generate the final reconstructed output.
The fixed filtering stage aims to provide an initial level of artifact suppression before online adaptation.
To achieve content adaptivity, ECM employs a fine-grained gradient-based classifier that assigns reconstructed samples to 7168 classes based on local directional and activity features.
These classes are mapped to a smaller set of fixed filters that are applied during decoding.
While effective, this design introduces computational complexity and substantial redundancy: a large number of classes are available, yet many classes exhibit similar local structures and ultimately share similar filtering behavior. 
Reducing the number of fixed filters without sacrificing filtering performance is therefore important for achieving low hardware complexity and improved scalability.

Prior studies have mainly focused on reducing the number of filter sets through online and encoder-side decisions. 
In VVC \cite{Karczewicz2021inloop, Bross2021OVVC}, filter sets come from two sources: (i) online-trained adaptation parameter sets (APS), which require the explicit signaling of filter coefficients, and (ii) offline-trained fixed-filter sets, where only the set index is transmitted.
To reduce the signaling burden, \cite{Fang2023Optimization} monitors the usage statistics of these two sources and adaptively disables the fixed-filter component if it is rarely selected for a given frame.
In addition, \cite{Huang2024fast} reduces the number of transmitted filter sets via class merging. 
This method groups classes with similar statistics after a first-pass classification, thereby reducing the number of distinct filters sent in the bitstream.
In contrast, our work targets redundancy in the set of fixed filters, enabling a smaller fixed-filter set without relying on online merging or on/off decisions.

Graph signal processing (GSP) has been successfully applied to image and video processing tasks \cite{cheung2018graph, wang2014graph, liu2017random}.
In these applications, local pixel neighborhoods are represented as graphs whose edge weights are determined by the intensity difference between connected pixels. Our work is motivated by the observation that, since structural information is already captured by the graph, a single graph filter on an input-dependent graph has the potential to mimic the filtering of multiple input-independent fixed filters. 
As an example, the popular input-dependent bilateral filter can be viewed as a single low-pass graph filter on an input-dependent graph \cite{gadde2013bilateral}.  
We expect GSP-based methods to naturally encode pixel-intensity relationships and local structural information in a manner invariant to certain geometric transformations, since graphs that differ only by geometric rotation are mathematically equivalent.
These properties suggest that some neighborhoods assigned to \textit{different fixed filters} by the gradient-based classifiers of ECM can be represented by similar graphs (e.g., rotated versions of the same graph) and can therefore be processed 
with a \textit{single graph filter}.
In this paper, we propose a graph-based fixed filtering framework for ALF, which 
enables multiple fixed-filter classes to share the same graph filters, 
resulting in a substantially reduced set of fixed filters with practically no impact in filtering performance.

\section{Preliminaries}
\label{sec:prelim}
\begin{figure}
    \centering
    \subfloat[\label{sfig:9x9FFcoeff}]{\includegraphics[width=0.4\linewidth]{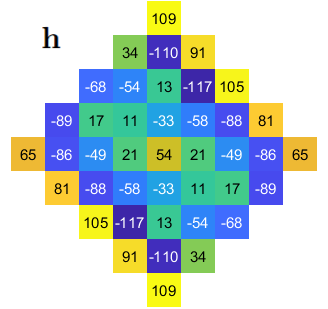}}
    \subfloat[\label{sfig:9x9graph}]{\includegraphics[width=0.4\linewidth]{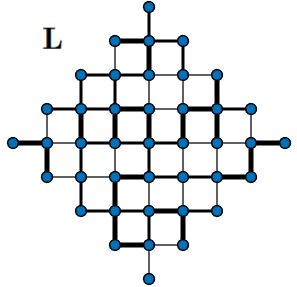}}
    \caption{Example fixed filtering operators on a $9\times9$ diamond neighborhood: (a) baseline fixed-filter coefficients $\mathbf{h}$ (color-coded and labeled), used in $\hat{y}=\mathbf{h}^{\top}\mathbf{x}$; (b) 4-connected weighted graph Laplacian $\mathbf{L}$ (thicker edges represent larger weights) used in $\hat{y}=\mathbf{s}^{\top}h(\mathbf{L})\mathbf{x}$.}

    \label{fig:toyExp9x9}
\end{figure}

Consider a weighted undirected graph $\mathcal{G} = (\mathcal{V}, \mathcal{E}, \mathbf{W})$, where $\mathcal{V}$ denotes a set of vertices that correspond to pixels in an image block, and $\mathcal{E}$ is a collection of edges describing the pairwise similarity between pixels. Denote the matrix entry $w_{i,j}$ in $\mathbf{W}$ as the weight associated with the edge $(i,j) \in \mathcal{E}$, and let $w_{i,j} = 0$ whenever $(i,j) \notin \mathcal{E}$.
The random-walk Laplacian is then given by
\begin{equation}
\mathbf{L} = \mathbf{I} - \mathbf{D}^{-1} \mathbf{W},
\label{eq:RW_L}
\end{equation}
where $\mathbf{D}$ is the diagonal degree matrix with each diagonal element $d_{i,i}$ equal to the sum $\sum_{j=1}^n w_{i,j}$.
The random-walk Laplacian provides a normalized representation of local signal variations on the graph and is widely adopted in graph-based filtering and restoration tasks
\cite{wang2014graph, liu2017random, Milanfar2013tour}.
Given an input $\mathbf{x} \in \mathbb{R}^{N}$, in our case, pixel values in a local neighborhood, defined on the nodes of a graph, 
a polynomial graph filter produces an output $\mathbf{y} \in \mathbb{R}^{N}$,
\begin{equation}
\mathbf{y} = h(\mathbf{L}) \mathbf{x} = \sum_{k=0}^{K} b_k \mathbf{L}^k \mathbf{x}.
\label{eq:graphFiltering}
\end{equation}
In fixed filtering for ALF, each pixel inherits a class label from its corresponding $2\times2$ block, based on fine-grained gradient-based classification of 7168 classes.
It is then filtered using a class-specific fixed filter defined over a larger local support, i.e., a $9\times9$ diamond-shaped neighborhood shown in Fig.~\ref{fig:toyExp9x9}(a).
The fixed-filter output is computed and retained independently for each pixel based on its assigned class, regardless of whether neighboring pixels share the same class label.
Letting $\mathbf{h}\in\mathbb{R}^{N}$ denote the vectorization of class-specific fixed-filter coefficients, the baseline fixed filtering computes the center-pixel output via the dot product $\hat{y}_{\text{FF}} = \mathbf{h}^\top \mathbf{x}.$

In the graph-based formulation, filtering follows the same per-pixel procedure.
A graph is constructed over a $9\times9$ local neighborhood shown in Fig.~\ref{fig:toyExp9x9}(b), and only the filter output corresponding to the center node is retained.
This can be expressed by introducing a selection vector $\mathbf{s}\in\mathbb{R}^{N}$, whose entries are zero except at the center position, yielding
\begin{equation}
\hat{y} = \mathbf{s}^\top h(\mathbf{L}) \mathbf{x}.
\end{equation}

\section{Graph-Based Fixed Filter Design for ALF}
\label{sec:GBFF}
This section presents the proposed graph-based fixed-filter design, including the problem formulation, graph construction strategy, and offline training procedure. 

\subsection{Problem Formulation}
\label{subsec:FFALF_ProbForm}

\begin{figure}
    \centering
    \includegraphics[width=0.95\linewidth]{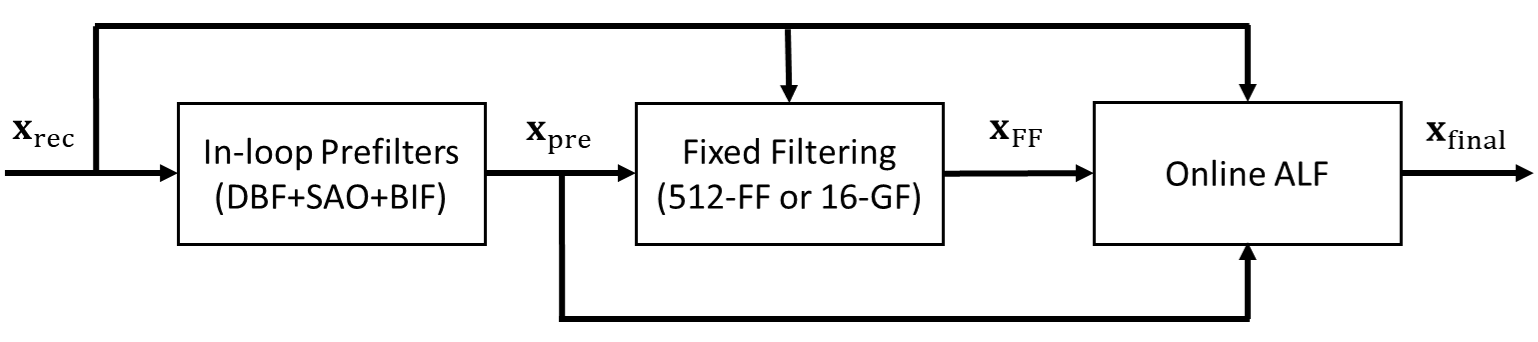}
    \caption{Block diagram of the ECM \cite{coban2025ecm19} ALF pipeline. We modify the fixed filter component by replacing 512 input-independent filters with 16 graph filters operating on local, input-dependent graphs.}
    \label{fig:blkdigm}
\end{figure}

ALF is employed as an in-loop post-processing stage to improve the quality of reconstructed samples by exploiting local spatial correlations.
As illustrated in Fig.~\ref{fig:blkdigm}, the ECM \cite{coban2025ecm19} ALF framework consists of three major stages: in-loop prefiltering, fixed filtering, and online ALF.
The in-loop prefiltering stage includes deblocking filtering (DBF), sample adaptive offset (SAO), and bilateral in-loop filtering (BIF). 
In this paper, we focus on the fixed filtering stage, which in ECM consists of 512 fixed filters (512-FF). We will propose methods to replace them with a much smaller set of 16 graph filters (16-GF).
Fixed filters constitute an important deterministic component of ALF.
They are trained offline and applied to reconstructed samples based on local structural characteristics.
In ECM, a gradient-based classifier assigns each $2\times2$ luma block sample to one of $56 \times 8 \times 16 = 7168$ classes,
corresponding to $56$ gradient direction categories characterized by horizontal/vertical and diagonal edge strengths $(E_{\text{HV}}, E_{\text{D}})$, 
$8$ $\L_2$-norm–based activity levels $A_{\text{L2}}$, and $16$ $\L_1$-norm–based activity levels $A_{\text{L1}}$.

\begin{figure} [t]
    \centering
    \subfloat[\label{sfig:class78rot45}]{\includegraphics[width=0.8\linewidth]{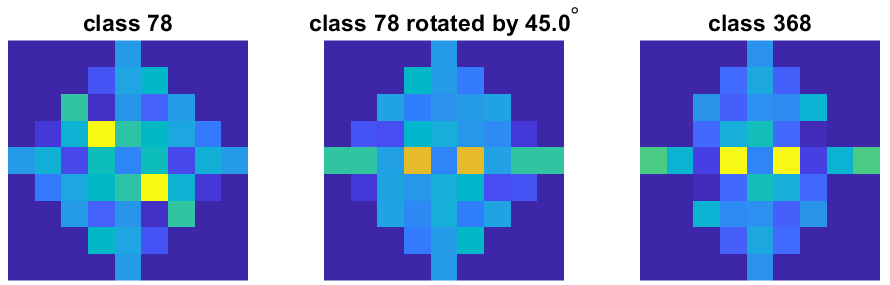}}
    \hspace{0.03\linewidth}
    \\ 
    \vspace{-8pt} 
    \subfloat[\label{sfig:class170rot15}]
    {\includegraphics[width=0.8\linewidth]{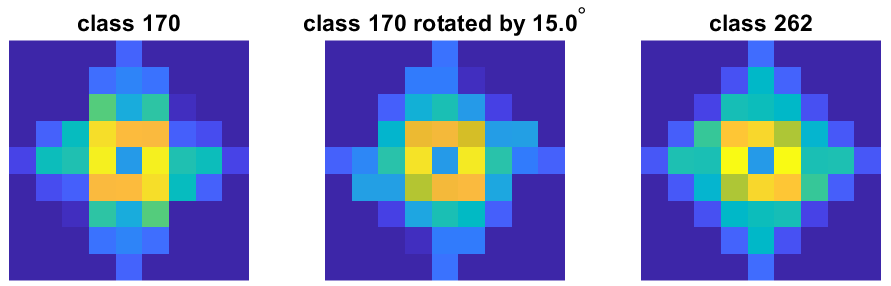}}
    \vspace{-4pt}
    \caption{Examples of rotation-based matching between 512 fixed-filter classes on the $9\times9$ diamond support. Colors encode coefficient values. Left: source class coefficients; middle: source class rotated by $\theta$ (interpolated on the diamond support); right: best-matching class at angle $\theta$. (a) $\theta=45^\circ$: class 78 $\rightarrow$ class 368. (b) $\theta=15^\circ$: class 170 $\rightarrow$ class 262.}
    
    \label{fig:FFRotateExp}
\end{figure}

Let $\mathbf{x} \in \mathbb{R}^{N}$ denote the reconstructed pixel intensities within a local neighborhood, and let $y$ denote the corresponding ground-truth intensity at the center pixel.
Fixed filtering estimates $\hat{y}$ from $\mathbf{x}$ using a class-dependent linear filter trained offline.
Although the classifier produces $7168$ distinct class indices, only $512$ fixed filters are trained and implemented. 
A predefined many-to-one mapping assigns multiple gradient-based classes to the same fixed filter, reflecting significant redundancy in the classifier space and enabling reduced storage and implementation complexity.
Moreover, the gradient-based classification and associated fixed filters are not fully rotation-invariant, as local structures related by geometric rotation may be assigned to different classes and processed using different filters.
As illustrated in Fig.~\ref{fig:FFRotateExp}, by applying an appropriate rotation to the $9\times9$ diamond support, some fixed filters can be revealed to be related: The two fixed filters of Fig.~\ref{fig:FFRotateExp}(a) and the two fixed filters of Fig.~\ref{fig:FFRotateExp}(b) could each be implemented with a single filter and a suitable rotation of the pixels in the diamond region. 

\begin{figure} [t]
    \centering
    \subfloat[\label{sfig:5nodesLeft}]{\includegraphics[width=0.47\linewidth]{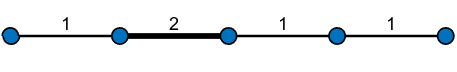}}
    \hspace{0.03\linewidth}
    \subfloat[\label{sfig:5nodesRight}]
    {\includegraphics[width=0.47\linewidth]{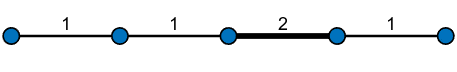}}
    \caption{1-D toy example showing graph isomorphism. A 5-pixel row is represented as a weighted path graph with unit edge weights, except one highlighted edge of weight $2$ at: (a) left of center; (b) right of center.}
    \label{fig:toyExp5nodes}
\end{figure}

To address this redundancy, a graph-based representation is introduced to further reduce the number of fixed filters.
Each local neighborhood is modeled as a graph signal, with pixel-intensity relationships encoded through a graph Laplacian constructed for each training sample.
As illustrated in Fig.~\ref{fig:toyExp5nodes}, neighborhoods that differ only by a geometric transformation (e.g., symmetric or rotations) can yield isomorphic graphs and Laplacians that are equivalent up to a node permutation.
This allows a single set of graph-based fixed-filter coefficients to be shared across multiple fixed-filter classes related by such transformation.

\subsection{Graph Construction Strategies}
\label{subsec:graphConstruct}


For each training sample, a graph is constructed on the $9\times9$ diamond-shaped local neighborhood centered at the target pixel, consistent with the support of the fixed filters.
Local graphs with either 4- or 8-connected neighborhoods are considered.
An example of a 4-connected graph is shown in Fig.~\ref{fig:toyExp9x9}(b), and the corresponding 8-connected graph is obtained by including the diagonal connections.
Given the fixed local connectivity, edge weights are defined solely in terms of pixel-intensity differences.
Specifically, the weight between two connected pixels $i$ and $j$ is given by
\begin{equation}
w_{ij} = \exp\!\left(-\frac{(x_i - x_j)^2}{\sigma_r^2}\right),
\end{equation}
where $x_i$ and $x_j$ denote the pixel intensities used for the graph construction.
This formulation captures local intensity similarity, while the geometric relationship between pixels is implicitly encoded by the fixed neighborhood connectivity.
The graph construction is adapted to the compression noise level by choosing the intensity-domain bandwidth $\sigma_r$  proportional to the quantization step, so  more aggressive filtering is used when larger quantization errors are expected~\cite{wennersten2017bilateral}.
Specifically, we define
\begin{equation}
\sigma_r = 0.35 \cdot Q_{\text{step}},
\end{equation}
where the quantization step size $Q_{\text{step}}$ is given by
\begin{equation}
Q_{\text{step}} = 2^{\frac{\text{QP}-4}{6}}.
\end{equation}
The corresponding random-walk Laplacian in \eqref{eq:RW_L} is used as the graph filtering operator in \eqref{eq:graphFiltering}. 
From an implementation perspective, to avoid redundant computation of edge weights for overlapping neighborhoods, a weighted graph can be constructed once for the entire decoded frame.
The local graph corresponding to each target pixel is then obtained by cropping the subgraph associated with its $9\times9$ diamond-shaped neighborhood.

The choice of graph connectivity influences both the expressiveness of the local graph representation and the polynomial order required by the graph filter. 
As will be shown in Section~\ref{subsec:numClus_PolyOrder}, the 8-connected graph configuration achieves a performance comparable to the 4-connected alternative while reaching its best performance at a lower polynomial order. 
This behavior suggests that increased local connectivity enables more efficient propagation of structural information through the graph Laplacian.

\begin{algorithm}[t]
\caption{Offline Training of Graph-Based Fixed Filters}
\label{alg:gbff}
\begin{algorithmic} [1]
\Require Training samples grouped into $512$ classes; polynomial order $K$; regularization $\lambda$; number of clusters $M$
\Ensure Cluster-level filters $\{\tilde{\mathbf{b}}_m\}_{m=1}^{M}$ and partitions $\{\mathcal{C}_m\}_{m=1}^{M}$

\For{$c = 1$ to $512$}
    \State Construct $\mathbf{L}^{(n)}$ for each sample $n$ in class $c$ 
    \State Estimate $\mathbf{b}_c$ via ridge regression (Sec.~3.3)
\EndFor

\State Compute clusters $\{\mathcal{C}_m\}_{m=1}^{M}$ by weighted $K$-means on $\{\mathbf{b}_c\}_{c=1}^{512}$

\For{$m = 1$ to $M$}
    \State Aggregate all samples from classes in $\mathcal{C}_m$
    \State Retrain coefficients $\tilde{\mathbf{b}}_m$ via ridge regression
\EndFor

\State \Return $\{\tilde{\mathbf{b}}_m\}_{m=1}^{M}$ and $\{\mathcal{C}_m\}_{m=1}^{M}$
\end{algorithmic}
\end{algorithm}

\subsection{Training of Graph-Based Fixed Filters}
\label{subsec:trainGBFF}

The full training procedure is summarized in Algorithm~\ref{alg:gbff}.
The proposed graph-based fixed filters are learned offline by exploiting the existing fixed-filter class structure in ECM.
Rather than learning filters directly from the full classifier space, the learning process starts from the $512$ fixed-filter classes and further reduces redundancy among them through coefficient-space clustering.

\subsubsection{Initial Per-Class Training}

In the first stage, a graph-based fixed filter is trained independently for each of the $512$ fixed-filter classes.
For a given class, let $\{(\mathbf{x}^{(n)}, \mathbf{L}^{(n)}, y^{(n)})\}_{n=1}^{N_c}$ denote its set of training samples, where $\mathbf{x}^{(n)}$ is the reconstructed input signal defined on the local neighborhood, $\mathbf{L}^{(n)}$ is the graph Laplacian constructed for that sample as described in Sec.~3.2, and $y^{(n)}$ is the corresponding clean target value.
Note that the graph Laplacian is constructed independently for each training sample, and no shared graph structure is assumed within a class.
For each class, the graph-based fixed filter is parameterized as a $K$-th order polynomial graph filter.
The filter coefficients $\mathbf{b} = [b_0,\ldots,b_K]^\top$ are obtained by solving a ridge regression problem,
\begin{equation}
\label{eq:ridgeRegression}
\min_{\mathbf{b}} 
\sum_{n=1}^{N_c}
\left(
y^{(n)} - \mathbf{s}^\top \sum_{k=0}^{K} b_k \big(\mathbf{L}^{(n)}\big)^k \mathbf{x}^{(n)}
\right)^2
+ \lambda \|\mathbf{b}\|_2^2,
\end{equation}
where $\mathbf{s}$ denotes the selection vector that extracts the center pixel from the graph filter output, and $\lambda$ is the regularization parameter, which 
we empirically set to  
$\lambda = 10^{-2}$ to prevent excessively large filter coefficients.
This procedure yields $512$ sets of graph filter coefficients, one for each of the ECM fixed-filter classes.

\subsubsection{Coefficient-Space Clustering}

To further reduce the number of graph filters, the $512$ learned graph filters are grouped using $K$-means clustering in the coefficient space.
Each graph filter is represented by its coefficient vector $\mathbf{b}$, and clustering is performed based on coefficient similarity.
This coefficient-space clustering reflects similarity in filtering behavior, enabling classes with similar local neighborhood structures to share a single graph-based fixed filter.
A weighted $K$-means algorithm is adopted to address class imbalance, with weights proportional to the number of training samples per class.
Formally, let $\mathbf{b}_i \in \mathbb{R}^{K+1}$ denote the graph filter coefficient vector learned for the $i$-th fixed-filter class, and let $\alpha_i$ denote its associated weight, defined as the number of training samples in that class.
Given a target number of clusters $M$, weighted $K$-means clustering seeks to partition the $512$ coefficient vectors into $M$ clusters by minimizing
\begin{equation}
\min_{\{\mathcal{C}_m\}, \, \{\boldsymbol{\mu}_m\}}
\sum_{m=1}^{M}
\sum_{i \in \mathcal{C}_m}
\alpha_i \,
\big\| \mathbf{b}_i - \boldsymbol{\mu}_m \big\|_2^2,
\end{equation}
where $\mathcal{C}_m$ denotes the index set of classes assigned to the $m$-th cluster and $\boldsymbol{\mu}_m$ is the corresponding cluster centroid.

\subsubsection{Cluster-Level Retraining}

Rather than directly averaging the graph filter coefficients within each cluster, a second training stage is performed to fully exploit the aggregated training data.
For each cluster, all training samples associated with the mapped classes are aggregated, and a cluster-level graph filter is learned by re-solving the ridge regression problem in \eqref{eq:ridgeRegression} using this aggregated training set.


\section{Design Analysis and Ablation Study}
\label{subsec:designAnalysis}

This section analyzes the impact of key design choices in the proposed graph-based fixed filtering framework.
As illustrated in Fig.~\ref{fig:blkdigm}, the ALF processing in ECM consists of multiple stages. 
The reconstructed signal $\mathbf{x}_{\text{rec}}$ is first processed by the in-loop prefilters, producing the prefiltered signal $\mathbf{x}_{\text{pre}}$. 
This signal is then fed to the fixed filtering stage, where the fixed filters are applied to generate $\mathbf{x}_{\text{FF}}$. 
Finally, the online ALF stage aggregates the outputs of the previous stages and applies Wiener filtering to generate the final reconstruction $\mathbf{x}_{\text{final}}$.
This multi-stage structure provides multiple candidate signals that can be used either for graph construction or as training targets, enabling different design choices. 

\subsection{Experimental Setup}
\label{subsec:expSetup}

Experiments are conducted on the BVI-DVC dataset \cite{ma2022BVIDVC}, which contains a diverse set of natural video sequences with varying spatial and temporal characteristics.
A total of $200$ sequences are used, with $180$ sequences for training and $20$ sequences reserved for testing.
Experiments are conducted using the $480\times272$ resolution, focusing on luma (Y) samples with 10-bit depth.
All experiments are performed at the quantization parameter $\text{QP}=37$, a typical operating point for evaluating compression artifacts.
Performance is measured using the mean absolute error (MAE) between the filtered and original clean samples.
For comparison, we report the error between the clean input and output at different stages, including (i) samples before fixed filtering $\mathbf{x}_{\text{pre}}$, (ii) fixed filtering output $\mathbf{x}_{\text{FF}}$, and (iii) ALF output $\mathbf{x}_{\text{final}}$ of the ECM baseline under the same experimental conditions.

\subsection{Graph Connectivity}
\label{subsec:graphConnectivity}

\begin{figure}
    \centering
    \subfloat[\label{sfig:compGraphConnect}]{\includegraphics[width=0.353\linewidth]{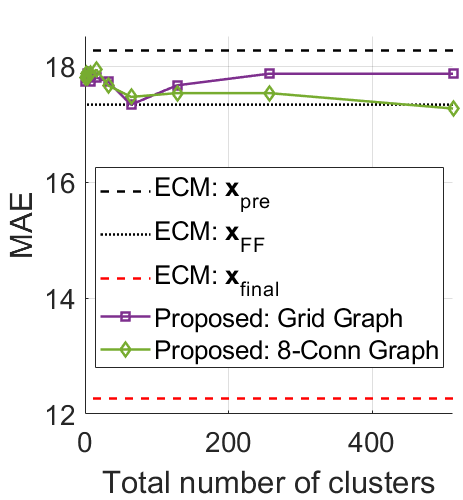}}
    \subfloat[\label{sfig:compGraphInput}]{\includegraphics[width=0.323\linewidth]{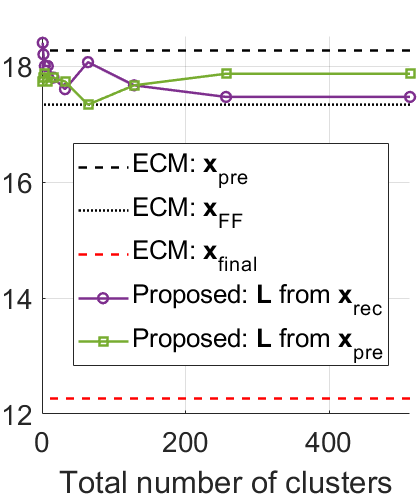}}
    \subfloat[\label{sfig:compTargetOutput}]{\includegraphics[width=0.323\linewidth]{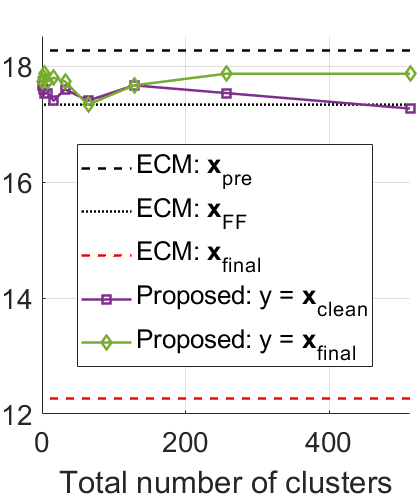}}
    \caption{Comparison of filtering performance in terms of MAE versus the total number of clusters under different configurations. 
    (a) Graph construction using either (i) 4-connectivity (grid graph) or (ii) 8-connectivity (8-conn graph). 
    (b) Graph edge weighted computed from sample (i) before in-loop prefiltering $\mathbf{x}_{\text{rec}}$ or (ii) after in-loop prefiltering $\mathbf{x}_{\text{pre}}$.
    (c) Target outputs for training selected to be (i) original clean samples $\mathbf{x}_{\text{clean}}$ or (ii) ALF outputs $\mathbf{x}_{\text{final}}$.}

    \label{fig:MAEvsNumClus}
\end{figure}
Fig.~\ref{fig:MAEvsNumClus} shows the relation between MAE and the number of graph-based fixed filters (clusters).
Fig.~\ref{fig:MAEvsNumClus}(a) compares the performance of 4-connected and 8-connected graph constructions across different number of clusters.
Although the 8-connected graph outperforms the 4-connected graph when the number of clusters is large, the two types of graphs perform similarly when the number of clusters is small, which is the more practical operating point. 

\subsection{Graph Construction Input and Training Target Selection}
Fig.~\ref{fig:MAEvsNumClus}(b) compares different choices of reconstructed samples used to build the graph Laplacian $\mathbf{L}$ in \eqref{eq:ridgeRegression}. 
We consider samples before in-loop prefiltering $\mathbf{x}_{\text{rec}}$ and after in-loop prefiltering $\mathbf{x}_{\text{pre}}$.
When the number of clusters is small, constructing graphs from the prefiltered samples $\mathbf{x}_{\text{pre}}$ consistently yields better performance.
Although using unfiltered samples $\mathbf{x}_{\text{rec}}$ may be beneficial when a large number of clusters is available, this advantage is less relevant to the goal of reducing the number of fixed filters in modern coding software.
Therefore, graphs constructed from $\mathbf{x}_{\text{pre}}$ are adopted in the proposed framework.

The choice of training target $y$ in \eqref{eq:ridgeRegression} is also examined in Fig.~\ref{fig:MAEvsNumClus}(c) by comparing two candidate signals: the original clean samples $\mathbf{x}_{\text{clean}}$ and the output of the online ALF stage $\mathbf{x}_{\text{final}}$.
Across all settings, using the clean samples as $y$ achieves the lowest MAE. 
Thus, clean samples provide the most appropriate supervision, which is adopted in the proposed framework.

\subsection{Number of Clusters and Polynomial Order}
\label{subsec:numClus_PolyOrder}

\begin{figure}
    \centering
    \subfloat[\label{sfig:compGraphPolyOrder_clus8}]{\includegraphics[width=0.45\linewidth]{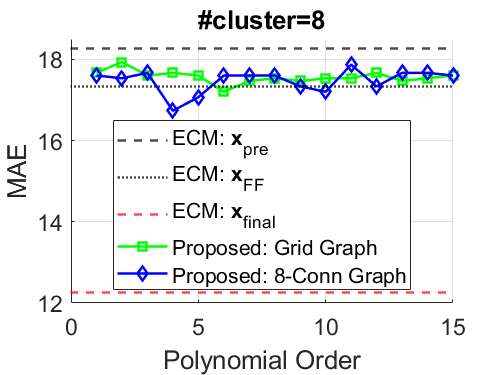}}
    \subfloat[\label{sfig:compGraphPolyOrder_clus16}]{\includegraphics[width=0.45\linewidth]{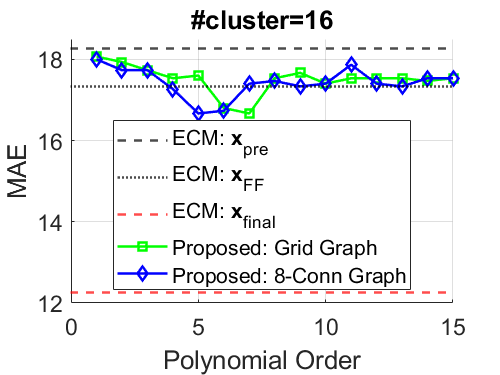}}  
    \\ \vspace{-8pt} 
    \subfloat[\label{sfig:compGraphPolyOrder_clus32}]{\includegraphics[width=0.45\linewidth]{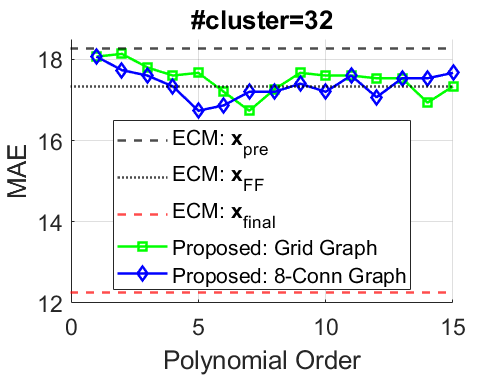}}
    \subfloat[\label{sfig:compGraphPolyOrder_clus64}]{\includegraphics[width=0.45\linewidth]{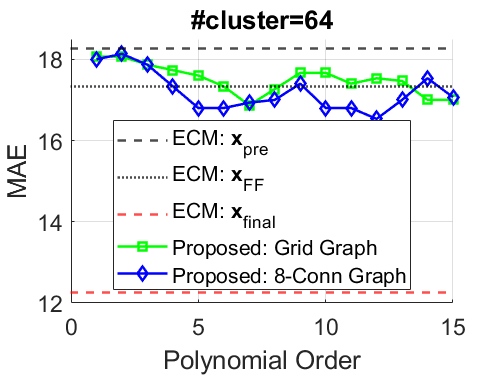}}
    
    \caption{Comparison of filtering performance in terms of MAE versus the filter order under different predefined number of clusters.}
    \label{fig:compPolyOrder_NumClusters}
\end{figure}

In this experiment, graphs are constructed from $\mathbf{x}_{\text{pre}}$ and clean samples are used as the target output for training.
Fig.~\ref{fig:compPolyOrder_NumClusters} shows the relation between MAE and the polynomial order of graph-based fixed filters.
The optimal polynomial order differs between the two graph connectivity configurations. 
8-connected graph achieves its best performance at order $K=5$, whereas the 4-connected graph requires a higher order $K=7$ to reach comparable performance.
This behavior indicates that richer local connectivity enables more efficient information propagation through the graph Laplacian, allowing a lower-order polynomial in the filter design.

Moreover, graph-based fixed filters can outperform the original fixed filters under appropriate configurations.
This suggests that many of the original $512$ fixed-filter classes exhibit similar filtering behavior and can be effectively represented by a compact set of graph-based filters.
Based on these findings, an 8-connected graph with $M=16$ clusters and polynomial order $K=5$ is selected as a favorable trade-off between performance and model complexity.

\subsection{Performance Comparison}
\label{subsec:perfcomp}
Based on the ablation analysis, we select a configuration that balances filtering performance and hardware complexity.
We use an \emph{8-connected} graph constructed from $\mathbf{x}_{\text{pre}}$, and train the filters to approximate $\mathbf{x}_{\text{clean}}$, with \(M=16\) clusters and polynomial order \(K=5\).
The performance of the recommended configuration is evaluated using five-fold cross-validation, with the corresponding standard deviations reported in Table~\ref{tab:maeComparison}.
The ECM fixed filtering (FF Output) achieves a relative MAE reduction of $5.0\%$, while the proposed graph-based fixed filtering (GF Output) achieves a larger reduction of $6.5\%$.
This improvement is achieved despite reducing the number of fixed filters from $512$ to $16$, while using the same gradient-based classifier for filter selection.
The worst performance is observed for the sequence \textit{WoodSJTU}, where the GF output increases the MAE from $28.57$ for the prefiltered input to $29.02$. 
This sequence contains dense, thin tree structures with strong backlighting and lens flare, where smoothing can easily introduce distortion.
To further evaluate generalization, we directly apply the GFs trained on the BVI-DVC dataset to the JCT-VC Class C dataset \cite{Bossen2013CommonTestConditions}, which consists of uncompressed sequences with a resolution of $832\times480$.
On this dataset, the MAE of the prefiltered input is $4.51$. The proposed GF output achieves an MAE of $4.38$, close to the ECM FF output MAE of $4.35$.
These results demonstrate that the proposed graph-based framework can more effectively suppress compression artifacts while using a substantially smaller set of fixed filters.

\begin{table}[t]
\centering
\caption{MAE Performance Comparison}
\label{tab:maeComparison}
\begin{tabular}{lccc}
\hline
 & Prefiltered Input & FF Output & GF Output \\
\hline
MAE & 22.34 & 21.23 & \textbf{20.89} \\
\hline
\makecell[l]{Relative \\ MAE Reduction} & 0\% & 5.0\% & \textbf{6.5\%} \\
\hline
\makecell[l]{Standard \\ Deviation} & 5.50 & 5.49 & 5.57 \\
\hline
\end{tabular}
\end{table}

\subsection{Structural Interpretation of GF Grouping}

\begin{figure}[t]
    \centering
    \includegraphics[width=0.99\linewidth]{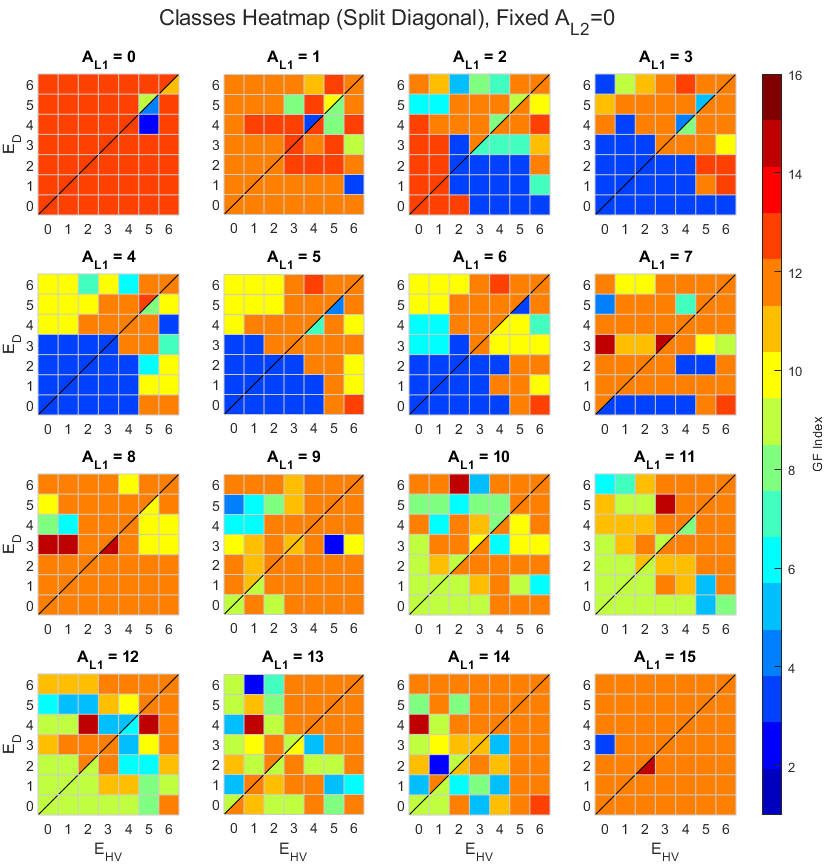}
    \caption{Mapping from ECM's gradient-based classifier classes to the learned graph filter (GF) indices. 
    Each cell corresponds to a single original gradient-based class, and its color denotes the assigned GF index. 
    The horizontal and vertical axes indicate the binned horizontal/vertical and diagonal edge strengths $(E_{\text{HV}}, E_{\text{D}})$. 
    Each subplot corresponds to a different $A_{\text{L1}}$ level (with $A_{\text{L2}}=0$).
    The diagonal split separates cases with $E_{\text{HV}}\geq E_{\text{D}}$ and $E_{\text{HV}} < E_{\text{D}}$.}
    \label{fig:GFHeatmap}
\end{figure}

To further understand the clustering behavior of the proposed graph filters, Fig.~\ref{fig:GFHeatmap} visualizes the mapping from the original fine-grained gradient-based classes to the learned GF indices. 
Overall, the learned GF combines classes that are similar in both orientation-related measures and activity levels. 
For instance, GF cluster~3 (blue) mainly occupies neighboring cells in the lower-left portion of the table and across several consecutive $A_{\text{L1}}$ levels. 
This suggests that the graph-based representation tends to group classes with structurally similar (i.e., approximately isomorphic) neighborhoods while reducing the number of distinct filters.

\section{Conclusion}
\label{sec:conclusion}

This paper presented a graph-based fixed filtering framework for ALF in ECM.
By modeling local neighborhoods as graph signals and learning fixed filters as polynomial graph filters, the proposed approach captures local structural information more effectively than gradient-based classification alone.
Exploiting structural similarity in the graph domain enables a compact representation of fixed filters, substantially reducing redundancy while preserving compatibility with the existing ALF framework.
Experimental results demonstrate that the proposed method improves fixed-filter performance compared to the ECM baseline while significantly reducing the number of required filters.
These results suggest that graph-based representations offer an effective and scalable solution for fixed filtering in adaptive loop filtering.


\bibliographystyle{IEEEbib}
\bibliography{strings,refs}

\end{document}